\documentclass[final,1p,times,twocolumn]{elsarticle}

\usepackage{amsmath,amssymb,amsthm,morefloats}
\usepackage{epsfig,psfrag,url}
\usepackage{graphicx}
\usepackage{verbatim}
\usepackage{subfigure}

\newcommand{\goto}{\rightarrow}
\newcommand{\bigo}{{\mathcal O}}

\def\XXint#1#2#3{{\setbox0=\hbox{$#1{#2#3}{\int}$}
     \vcenter{\hbox{$#2#3$}}\kern-.5\wd0}}

\DeclareMathOperator{\sech}{sech}

\newenvironment{mat}{\left[\begin{array}{ccccccccccccccc}}{\end{array}\right]}
\newcommand\bcm{\begin{mat}}
\newcommand\ecm{\end{mat}}

\newcommand{\bea}{\begin{eqnarray}}
\newcommand{\eea}{\end{eqnarray}}
\newcommand{\bean}{\begin{eqnarray*}}
\newcommand{\eean}{\end{eqnarray*}}
\newcommand{\ba}{\begin{array}}
\newcommand{\ea}{\end{array}}
\newcommand{\beqs}{\begin{equation*}\begin{split}}

\newtheorem{theorem}{Theorem}[section]

\journal{Physics Letters A}

\begin{document}

\begin{frontmatter}



\title{Dispersive and soliton perturbations of finite-genus solutions of the KdV equation: computational results}


\author{Thomas Trogdon and Bernard Deconinck}

\address{
Department of Applied Mathematics\\
 University of Washington\\
Campus Box 352420\\
 Seattle, WA, 98195, USA \\}

\begin{abstract}
All solutions of the Korteweg -- de Vries equation that are bounded on the real line are physically relevant, depending on the application area of interest. Usually, both analytical and numerical approaches consider solution profiles that are either spatially localized or (quasi)periodic. In this paper, we discuss a class of solutions that is a nonlinear superposition of these two cases: their asymptotic state for large $|x|$ is (quasi)periodic, but they may contain solitons, with or without dispersive tails. Such scenarios might occur in the case of localized perturbations of previously present sea swell, for instance. Such solutions have been discussed from an analytical point of view only recently. We numerically demonstrate different features of these solutions.
\end{abstract}

\begin{keyword}


KdV equation \sep soliton \sep dispersion \sep finite genus \sep Riemann-Hilbert problem \sep numerical computation

\end{keyword}

\end{frontmatter}


\section{Introduction}
\label{Section:Introduction}

The Korteweg -- de Vries (KdV) equation is one of the most studied nonlinear partial differential equations. It can be written as

\begin{align}\label{kdv}
q_t + 6qq_x + q_{xxx}=0,
\end{align}

\noindent where $x$ and $t$ represent a scaled spatial and temporal independent variable, respectively, and $q(x,t)$ is the function to be determined. The KdV equation arises in the study of long waves in shallow water, ion-acoustic waves in plasmas, and in general, describes the slow evolution of long waves in dispersive media \cite{ablowitz-segur-book}. The importance of the equation partially derives from how well its solutions match experimental observations \cite{Hammack1,Hammack2,Hammack3}. When comparing solutions with both experimental and numerical data, the so-called soliton \cite{ablowitz-segur-book} solutions have often taken a dominant place. Although these solutions are beyond doubt significant, their simple functional form and straightforward dynamical behavior have contributed to their prominence. Recent work on the numerical evaluation of solutions other than solitons has allowed for the detailed study of other important classes of solutions, such as dispersive tails \cite{TrogdonSOKdV} or (quasi-) periodic multi-phase solutions, the so-called finite-genus solutions. Although the finite-genus solutions have been computed before by others \cite{Deconinck-theta,klein,Lax}, their computation within the framework of Riemann-Hilbert problems \cite{TrogdonFiniteGenus} now allows for the investigation of the nonlinear superposition of such solutions with solitons and dispersive tails. The corresponding numerical method was discussed in detail in \cite{TrogdonDressing}. The analysis of such superposition solutions was examined by Egorova, Grunert and Teschl \cite{teschl-finite-gap} and by Mikikits-Leitner \& Teschl \cite{teschl-asymptotics}, including some scenarios that are not covered in \cite{TrogdonDressing}.

From an analysis point of view, the nonlinear superposition of finite-genus solutions and localized profiles containing dispersion and solitons requires the use of non-standard function spaces because of the asymptotic behavior of the finite-genus solutions, which are typically quasi periodic. This is not required in the study of the finite-genus solutions on their own, as these are finite-dimensional solutions. Only when the number of phases is allowed to increase without bound are function-theoretic considerations relevant \cite{McKean-Trubowitz}. In that case, one often restricts to periodic boundary conditions, allowing the use of standard spaces. Dispersion, on the other hand, is inherently infinite dimensional. Both solitons and dispersive tails fit within the framework of the inverse scattering transform \cite{TrogdonSOKdV, GGKM}, resulting in standard spaces with initial data that are localized on the whole line. When the solution contains both finite-genus components and dispersion with solitons, neither the use of standard spaces with periodic functions or with localized functions is possible. Egorova {\em et al.} \cite{teschl-finite-gap} introduce new spaces where the quasi-periodic behavior at both $x\rightarrow +\infty$ and $x\rightarrow -\infty$ is subtracted off, so that only localized functions remain. Note that the quasi-periodic behavior on either end is distinct, as is demonstrated numerically below.

The same issues that complicate the analysis prevent the numerical computation of the solutions using traditional methods: not only is the quasi-periodic behavior as $|x|\rightarrow \infty$ problematic, the presence of dispersion with its small but effectively faster (as $t$ increases) oscillations dooms any approach using a traditional time-stepping algorithm. A discussion of this is found in \cite{TrogdonSOKdV}. Our method avoids these issues completely: using the integrable nature of (\ref{kdv}) the superposition solutions are evaluated at any $x$ or $t$ by solving a Riemann-Hilbert problem, as briefly outlined below. The crux of the present work and that in \cite{TrogdonDressing} is that the approach of \cite{TrogdonSOKdV} to make the inverse scattering transform effective may be combined with that in \cite{TrogdonFiniteGenus}, where finite-genus solutions are computed, using the Riemann-Hilbert approach. The result is an effective numerical algorithm to compute the superposition solutions, which are in effect localized perturbations of finite-genus solutions.

The goal of this short paper is the computational investigation of some of the properties of the superposition solutions, especially with regards to the qualitative and quantitative differences that occur as a consequence of the nonlinear superposition. The discussion differs from that in \cite{TrogdonDressing} where the numerical method is discussed and presented in great detail. In contrast, only limited space is devoted to the dressing method \cite{FokasUnified} and to Riemann-Hilbert problems, while no mentioning is made of the numerical method, other than presenting some of its results.

\section{The Dressing Method}

The numerical method we sketch here is derived from the \emph{dressing method} \cite{FokasUnified,Dressing,ZakharovDressing}.  Before discussing the method, we discuss an analogous method applied to the linear equation
\begin{align}\label{linear-kdv}
q_t + q_{xxx} = 0.
\end{align}
One can write down a solution in Ehrenpreis form
\begin{align*}
q(x,t) = \int_\Gamma f(k) e^{ikx+ik^3t} dk,
\end{align*}
where $\Gamma$ is chosen so that the integral is convergent.  The solution of the initial-value problem on the line with decaying initial data is solved with:
\begin{align*}
f(k) = \frac{1}{2\pi} \int_\mathbb R e^{-ikx}q(x,0) dx, ~~ \Gamma = \mathbb R.
\end{align*}
The solution of the periodic problem, often solved in terms of a Fourier series, can be expressed in this form \cite{TrogdonPeriodic}.  Taking a constructive approach, we may specify both $f(k)$ and $\Gamma$ and construct a solution of \eqref{linear-kdv}.  This is the so-called direct method. An example solution is simply
\begin{align*}
q(x,t) = \int_{\Gamma} e^{ikx+ik^3t} dk,
\end{align*}
where $\Gamma$ is a contour in the upper-half plane such that $k^3t$ is real valued. This is directly related to the Airy function \cite{DLMF}.

To motivate the extension to nonlinear problems, we show the connection between contour integrals and \emph{Riemann--Hilbert problems}.  Given an oriented contour $\Gamma \subset \mathbb C$ and $G: \Gamma \goto \mathbb C^{2 \times 2}$ a Riemann--Hilbert problem (RHP) poses the task of finding a function $\Phi: \mathbb C \setminus \Gamma \goto \mathbb C^{m \times 2}$ that is analytic in $\mathbb C \setminus \Gamma$ and satisfies
\begin{align*}
\Phi^+(k) = \Phi^-(k) G(k),~~~ k \in \Gamma,
\end{align*}
where we use $\Phi(\infty) = I$ if $m=2$ (matrix RHP) and $\Phi(\infty) = [1,1]$ if $m=1$ (vector RHP).

Consider the RHP
\begin{align*}
\Phi^+(k) = \Phi^-(k)\begin{mat} 1 & 0 \\ e^{ikx+ik^3t}f(k) & 1 \end{mat}, ~~~ k \in \Gamma,~~~ \Phi(\infty) = I.
\end{align*}
Assuming sufficient decay of $f(k)$ as $|k| \goto \infty$ it can be shown that \cite{FokasComplexVariables}
\begin{align*}
\Phi(k) = \begin{mat} 1 & 0 \\ F(k) & 1 \end{mat}, ~~ F(k) = \frac{1}{2 \pi i} \int_{\Gamma} \frac{e^{iks+is^2t}f(s)}{s-k} ds.
\end{align*}
Then
\begin{align*}
- 2 \pi i \lim_{|k|\goto \infty} k \Phi_{21}(k) = \int_{\Gamma} e^{ikx+ik^3t}f(k) dk
\end{align*}
is a solution of \eqref{linear-kdv}.  The dressing method \cite{FokasUnified} is a generalization of this procedure to nonlinear PDEs.  Define the Pauli matrices
\begin{align*}
\sigma_3 = \begin{mat} 1 & 0 \\ 0 & -1 \end{mat}, ~~~ \sigma_2 = \begin{mat} 0 & 1 \\ 1 & 0 \end{mat}.
\end{align*}

We state the dressing method as a theorem.

\begin{theorem}\label{Theorem:Dressing}
Let $\Phi(k)$ solve the RHP
\begin{align*}
\Phi^+(k) = \Phi^-(k) e^{-\theta(k)\sigma_3} V(k) e^{\theta(k) \sigma_3}, ~~ k \in \Gamma,~~ \theta(k) = ikx+4ik^3t, ~~\Phi(\infty) = [1,1],
\end{align*}
where $\bar \Gamma = \Gamma$ (with orientation), $\det V(k) = 1$, $\overline{V(\bar k)} = V(-k)$ and $ V^{-1}(k) = \sigma_2 \overline{V(\bar k)} \sigma_2$.  Assume that the RHP has a unique solution that is sufficiently differentiable in $x$ and $t$ and all existing derivatives are $\bigo(1/k)$ as $k \goto \infty$.  Define
\begin{align}\label{Q-reconstruct}
\begin{mat} Q(x,t) & Q(x,t) \end{mat} = 2i \lim_{k \goto \infty} k \partial_x \Phi(k) \sigma_3.
\end{align}
Then $\Phi(k)$ solves
\begin{align}\begin{split}\label{jost-pair}
-\Phi_{xx} + 2ik \Phi_x \sigma_3 &- Q(x,t) \Phi = 0,\\
- \Phi_t +  4ik^3 \Phi \sigma_3 &= (2 Q(x,t)-4k^2)\left( \Phi_x - ik \Phi \sigma_3 \right) - Q_x(x,t) \Phi,
\end{split}\end{align}
and $Q(x,t)$ solves \eqref{kdv}.
\end{theorem}

\begin{figure}[t]
\centering
\includegraphics[width=.9\linewidth]{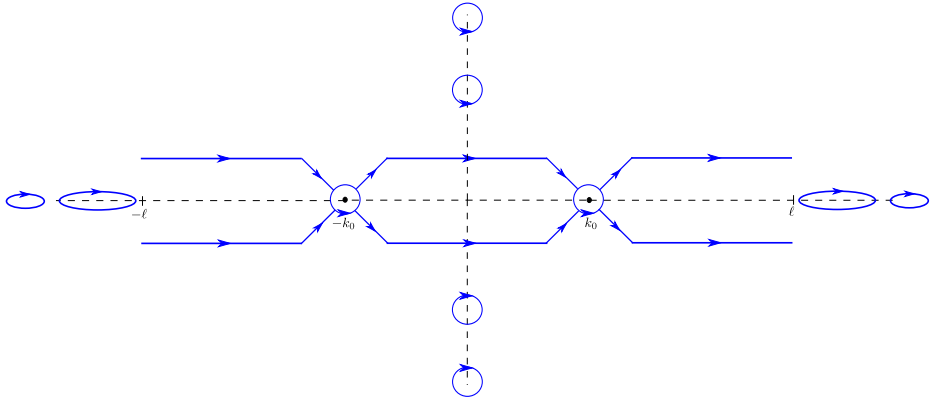}
\caption{\label{fg-RHP}
The jump contours of the RHP associated with the nonlinear superposition of a dispersive solution of \eqref{kdv} and a quasi-periodic solution of \eqref{kdv}. Here $\pm k_0$ are the stationary points of $\exp {\theta(k)}$ and $\ell > 0$ is a constant chosen for numerical purposes.
}
\end{figure}

Using these ideas a RHP was derived in \cite{TrogdonDressing} that corresponds to the nonlinear superposition of a quasi-periodic, finite-genus solution of \eqref{kdv} with a solution of the Cauchy initial-value problem on the line with rapidly decaying initial data.  The details of the RHP can be found in \cite{TrogdonDressing} and the jump contours are displayed in Figure~\ref{fg-RHP}.  The contours on the imaginary axis correspond to solitons in the solution.  The ellipses on the real axis correspond to a quasi-periodic background solution and the array of contours passing around $\pm k_0$ represent the dispersive aspects of the solution, see \cite{TrogdonSOKdV}.  As we discuss in the following section, no other numerical methods exist for computing such solutions.

\section{Numerical Results}

In this section we present the results of our computations.  We do not address accuracy.  An in-depth analysis can be found in \cite{TrogdonSOKdV,TrogdonFiniteGenus, TrogdonSONLS,TrogdonSONNSD}.  An overview of the numerical method for RHPs that we use is found in \cite{SORHFramework,SOPainleveII}.  These works address every case we consider here.  In general, the method takes in $x$ and $t$ as parameters and returns an approximation of the solution at the point $(x,t)$.  No time-stepping or spatial discretization is used to obtain the plots we display below.

\subsection{Genus two with two solitons}\label{Section:DSGT}

\begin{figure}[tbp]
\centering
\subfigure[]{\includegraphics[width=\linewidth]{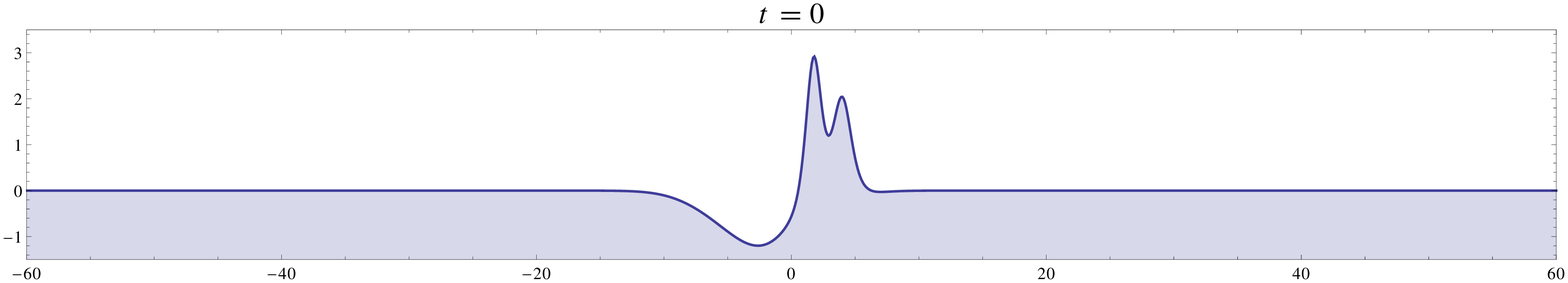} \label{DSGT-noG-0}}
\subfigure[]{\includegraphics[width=\linewidth]{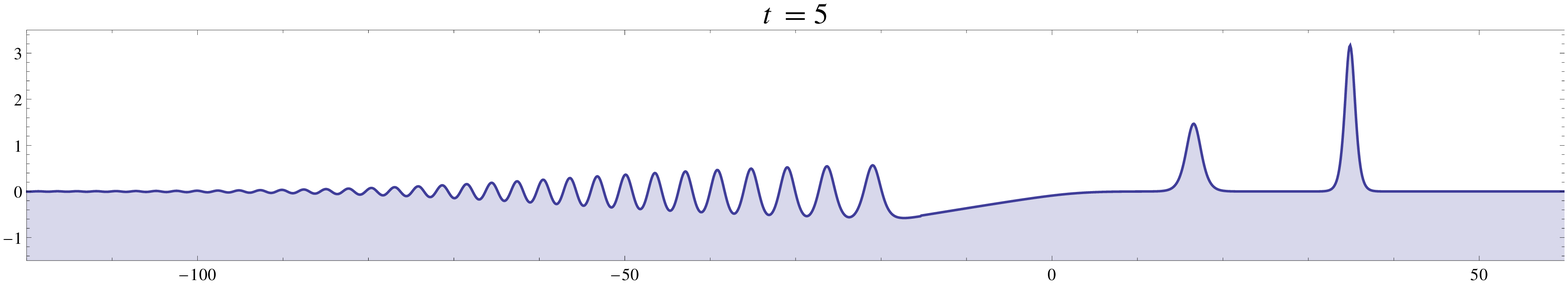} \label{DSGT-noG-5}}
\subfigure[]{\includegraphics[width=\linewidth]{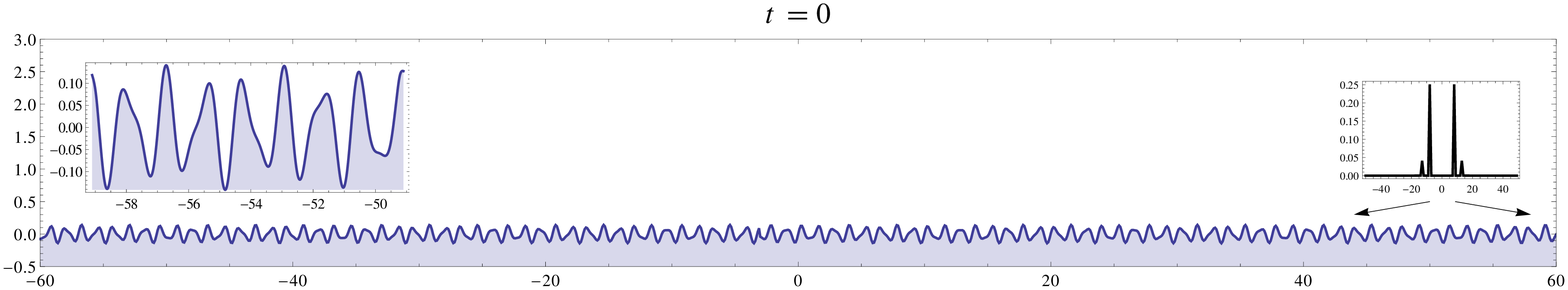} \label{DSGT-justG-0}}
\caption{(a) A dispersive solution of the KdV equation with two solitons at $t=0$. (b) A dispersive solution of the KdV equation with two solitons at $t = 5$. (c) A genus-two solution of the KdV equation. \label{DSGT-no-just}}
\end{figure}

\begin{figure}[tbp]
\centering
\subfigure[]{\includegraphics[width=\linewidth]{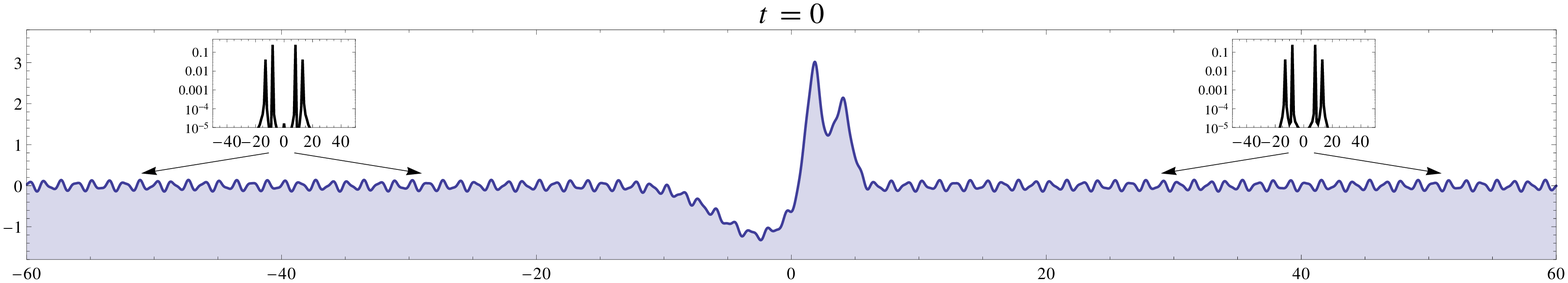}}
\subfigure[]{\includegraphics[width=\linewidth]{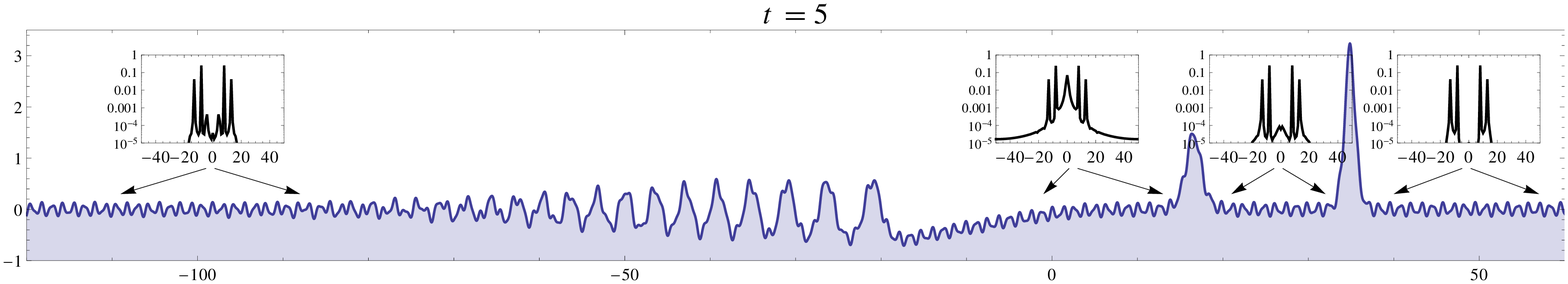}}
\caption{A two-soliton solution of the KdV equation with a quasi-periodic genus two background.  The insets show the amplitude of the DFT of the solution over the indicated regions to demonstrate the fundamental frequencies that are present. (a) The solution at $t=0$. (b) The solution at $t=5$. \label{DSGT}}
\end{figure}

The exact arrangement of contours shown in Figure~\ref{fg-RHP} produces the solution of the KdV equation shown in Figure~\ref{DSGT}.   We see significant dispersion interacting with the quasi-periodic background.  We emphasize that due to this quasi-periodic background, no other existing numerical methods can compute this solution.  Furthermore, the dispersion brings with it other numerical issues, see \cite{TrogdonSOKdV}.  We also display the solutions before superposition in Figure~\ref{DSGT-no-just}.  Removing the ellipses from Figure~\ref{DSGT} we obtain the (localized) solution in Figures~\ref{DSGT-noG-0} and \ref{DSGT-noG-5}.  Leaving only the ellipses in Figure~\ref{DSGT} we obtain the solution in Figure~\ref{DSGT-justG-0}.

The quasi-periodic background shown in Figure~\ref{DSGT} has four asymptotic regions: (i) $x \ll 0$, (ii) between the dispersive tail and the first soliton, (iii) between the two solitons, and (iv) $x \gg 0$.  Since KdV solitons always separate we expect $n+2$ regions when $n$ solitons are present in the solution.  We analyze these regions in more detail on a case-by-case basis in what follows.  In this paper we restrict to a genus two (or lower) background but we note there is no barrier preventing computations with higher genus background.

We compare our computed solutions with the analysis in \cite{teschl-finite-gap,teschl-asymptotics}.  In particular, we compute the amplitude of the discrete Fourier transform (DFT) of the solution in each of the regions outlined in Section~\ref{Section:DSGT}.  See Figure~\ref{DSGT} for these results.  The figure demonstrates that each region consists asymptotically of a solution of the KdV equation with the same two fundamental frequencies --- a genus two solution.   Furthermore, the results shown below in \eqref{DGO-Diff} indicates that this must be the same solution with phase shifts.  This can be seen rigorously through the analysis of the RHP \cite{teschl-asymptotics}.

\subsection{Genus two with four solitons}

\begin{figure}[t]
\centering
\subfigure[]{\includegraphics[width=\linewidth]{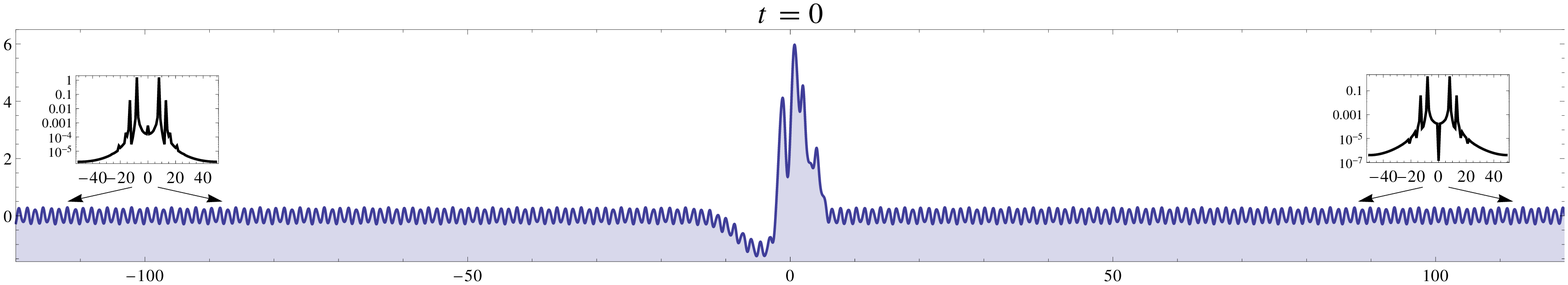}}
\subfigure[]{\includegraphics[width=\linewidth]{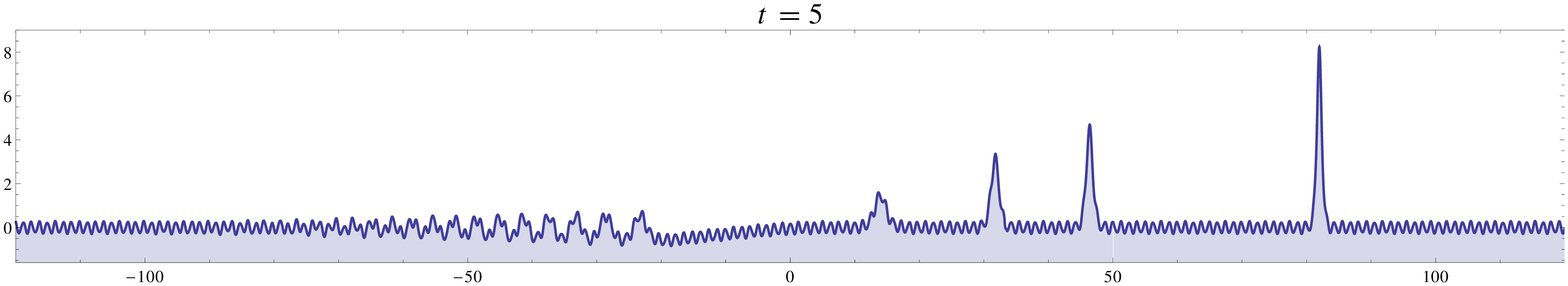}}
\caption{A dispersive four-soliton solution of the KdV equation with a quasi-periodic genus two background.  The insets show the amplitude of the DFT of the solution over the indicated regions to demonstrate the fundamental frequencies that are present.  (a) The solution at $t=0$. (b) The solution at $t=5$. \label{DSGT-4}}
\end{figure}

To further demonstrate the method we plot the evolution of a dispersive four-soliton solution on a small amplitude quasi-periodic background.  The quasi-periodicity of the background is not easily seen by the eye but the amplitude of the DFT clearly shows two fundamental frequencies.  This solution is shown in Figure~\ref{DSGT-4}.

\section{The Nonlocality of Superposition}

\begin{figure}[t]
\centering
\subfigure[]{\includegraphics[width=\linewidth]{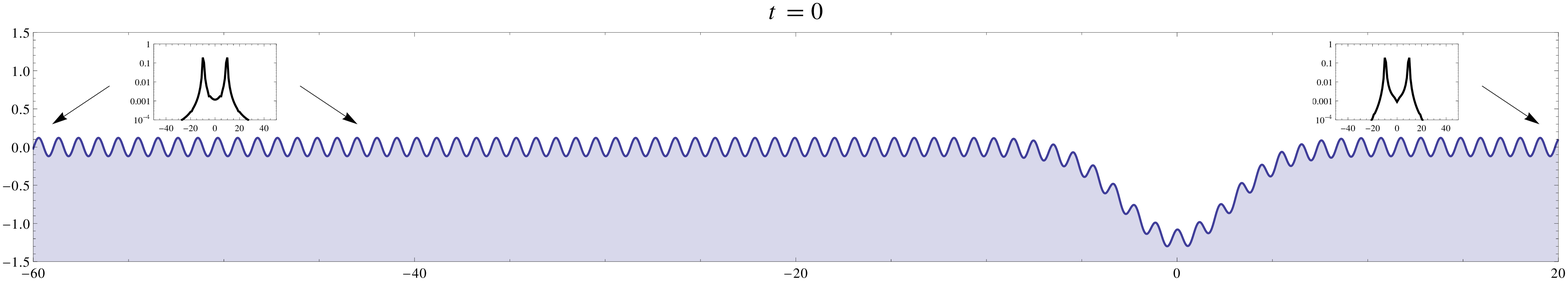}}
\subfigure[]{\includegraphics[width=\linewidth]{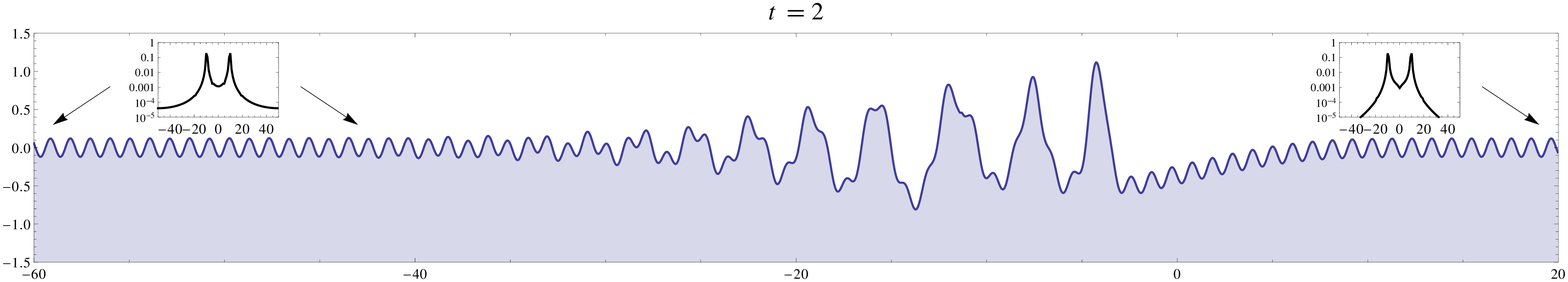}}
\subfigure[]{\includegraphics[width=\linewidth]{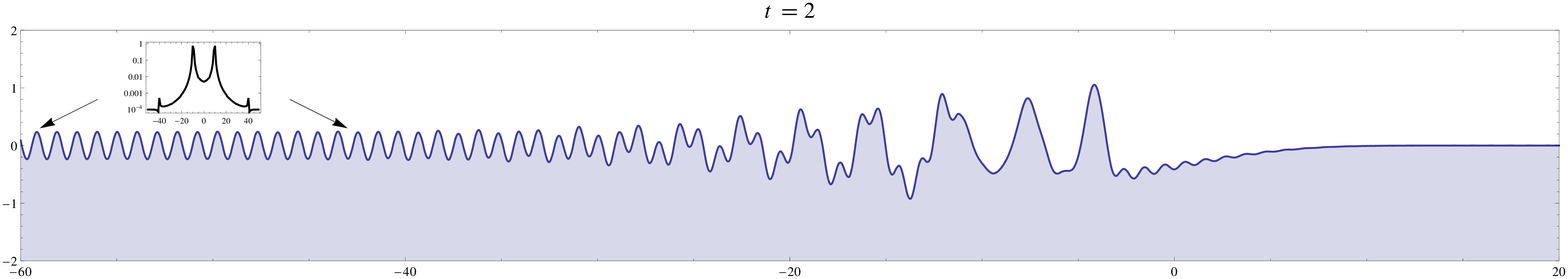} \label{DGO-Diff}}
\caption{A solution of the KdV equation with a periodic genus one background.  The insets show the amplitude of the DFT of the solution over the indicated regions to demonstrate the fundamental frequencies that are present.  (a) The solution at $t=0$. (b) The solution at $t=2$. (c) A plot of $q_2(x,2)-q_1(x,2)$.  This difference does not decay at infinity indicating that the nonlinear superposition of $q_0$ and $q_1$ is non-local. \label{DGO}}
\end{figure}

We may compute simpler solutions as well.  In Figure~\ref{DGO} we display a purely dispersive solution of \eqref{kdv} superimposed on a periodic genus one background.  Let $q_2(x,t)$ be this solution.  If we remove the dispersive contours from the RHP we obtain a periodic, genus one solution of the KdV equation.  Label this solution $q_1(x,t)$.  On the other hand, if we remove the ellipses on the real axis we obtain a purely dispersive solution of the KdV equation that decays at infinity. Label this solution $q_0(x,t)$. We measure the effect of nonlinearity by examining $q_2(x,t) - q_1(x,t)$. If the effect of nonlinearity was local then  $q_2(x,t) - q_1(x,t)$ would be a local perturbation of $q_0(x,t)$ and would decay at infinity.  In Figure~\ref{DGO-Diff} we see that this effect is non-local: for $x \ll 0$ we see that $q_2(x,t)$ is a phase-shifted genus one solution while for $x \gg 0$, $q_2(x,t) \sim q_1(x,t)$. This significantly complicates the computation of such solutions with classical methods because the solution cannot be approximated by a periodic solution.

\section{Nonlinear Peak Amplitude}

It is a relevant physical question to ask how the (quasi-)periodic background affects the soliton amplitude.  The method presented here allows this question to be examined for the KdV equation.  We remove dispersion from the solution.  The associated RHP contains only circles on the imaginary axis and ellipses on the real line.  Let $q_{\text{full}}(x,t)$ be the solution of the KdV equation found by solving this RHP.

In the absence of the ellipses, the RHP can be solved explicitly and the solution separates asymptotically into a sum of the form
\begin{align*}
q_{\text{s}}(x,t) \sim \sum_{j}^m 2 \kappa_j^2 \sech^2(\kappa_j(x- 4 \kappa_j^2t + \phi_j)) + \bigo(t^{-n}) \text{ for all } n > 0 \text{ as } t \goto \infty,
\end{align*}
with $\kappa_j < \kappa_{j+1}$, see Figure~\ref{DSGT-noG-5} for an example.  Therefore we know the maximum value of the solution for large time is $2 \kappa_m^2$.  Additionally, we can compute an interval $I_j(t)$ such that the peak of the $j$th soliton lies in $I_j(t)$.  We perform our comparison for the largest and therefore fastest moving soliton. Let $q_{\text{fg}}(x,t)$ be the finite-genus solution of the KdV equation found by removing the circles on the imaginary axis from the RHP.

\begin{figure}[t]
\centering
\includegraphics[width=.6\linewidth]{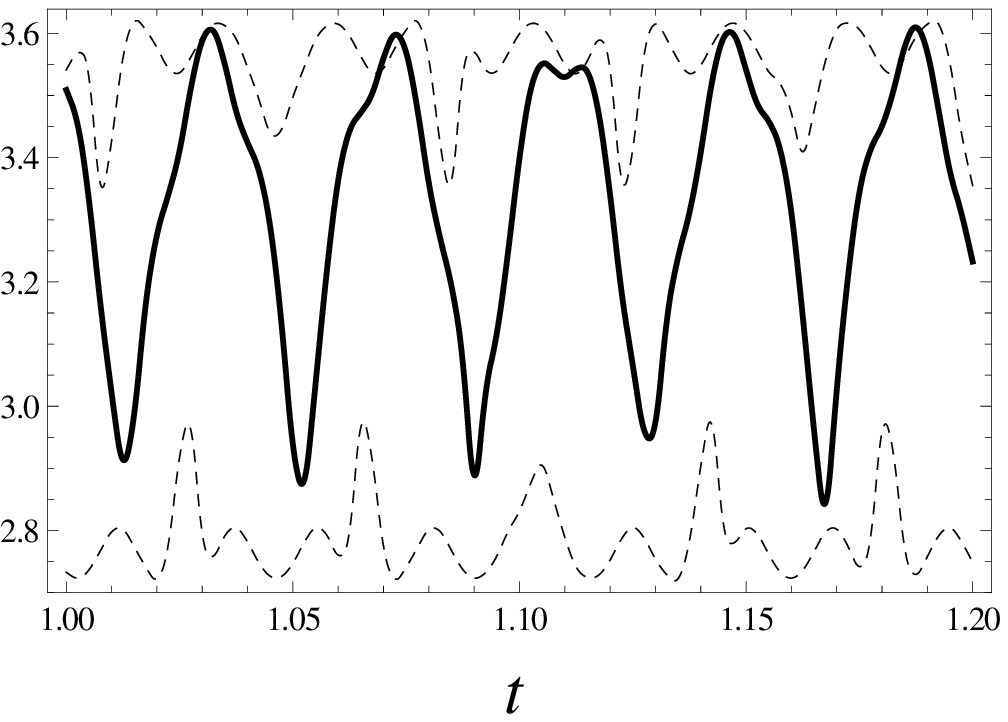}
\caption{\label{Nonlinear-Amp}
A plot of $\max_{x \in I_m(t) } q_{\text{full}}(x,t)$ (solid), $\max_{x \in I_m(t)} q_{\text{fg}}(x,t)+ 2 \kappa_m^2$ (dashed) and $\min_{x \in I_m(t)} q_{\text{fg}}(x,t)+ 2 \kappa_m^2$ (dashed) as functions of $t$.  For most times the dashed curves provide an upper and lower bound for $\max_{x \in I_m(t) } q_{\text{full}}(x,t)$ although the lower bound is not tight.}
\end{figure}

If, asymptotically, solutions superimpose linearly we would expect the maximum value $\max_{x \in I_m(t) } q_{\text{full}}(x,t)$ of the solution  to be bounded above by $\max_{x \in I_m(t)} q_{\text{fg}}(x,t)+ 2 \kappa_m^2$ and below by $\min_{x \in I_m(t)} q_{\text{fg}}(x,t) + 2 \kappa_m^2$.  We take $I_m(t)$ to be an interval of length $0.8$ and plot the above quantities in the two soliton, genus two case in Figure~\ref{Nonlinear-Amp}.  We see that $\max_{x \in I_m(t)} q_{\text{fg}}(x,t)+ 2 \kappa_m^2$ is not the upper bound due to nonlinear interactions although it is a fairly accurate upper estimate.  Importantly, we see that $\min_{x \in I_m(t)} q_{\text{fg}}(x,t)+ 2 \kappa_m^2$ seems to be a true lower bound but it does not closely track the solution.  This experiment indicates that a quasi-periodic background rarely amplifies the peak above maximum linear level and that it never suppresses the peak below the minimum linear level.

\section{Conclusions}

We have combined the dressing method and Riemann--Hilbert contour deformations with a numerical method for RHPs to compute a class of step-like finite-genus solutions of the KdV equation \cite{teschl-finite-gap} which we call superposition solutions.  Due to either quasi-periodicity (Figure~\ref{DSGT}) or the induced phase shift (Figure~\ref{DGO}) no other existing numerical methods can compute these solutions.  The KdV equation is one of the most studied nonlinear PDEs in the last half century, yet we are able to produce plots of physically relevant solutions that have not been seen before.

We are also able to examine the effect of a quasi-periodic background on the amplitude of solitons.  When compared with a linear scenario we see mild amplification over small time intervals.




\section*{Acknowledgments}

We acknowledge the National Science Foundation for its generous support through grant NSF-DMS-1008001 (BD,TT).  Any opinions, findings, and conclusions or recommendations expressed in this material are those of the authors and do not necessarily reflect the views of the funding sources.


\bibliographystyle{elsarticle-num}
\bibliography{PerturbedFG}

\def\cydot{\leavevmode\raise.4ex\hbox{.}}
  \def\cydot{\leavevmode\raise.4ex\hbox{.}} \def\cprime{$'$}
\begin{thebibliography}{10}
\expandafter\ifx\csname url\endcsname\relax
  \def\url#1{\texttt{#1}}\fi
\expandafter\ifx\csname urlprefix\endcsname\relax\def\urlprefix{URL }\fi
\expandafter\ifx\csname href\endcsname\relax
  \def\href#1#2{#2} \def\path#1{#1}\fi

\bibitem{ablowitz-segur-book}
M.~Ablowitz, H.~Segur, Solitons and the Inverse Scattering Transform, SIAM,
  Philadelpha, PA, 1981.

\bibitem{Hammack1}
H.~Segur, The {K}orteweg-de {V}ries equation and water waves. {S}olutions of
  the equation. {P}art 1, Journal of Fluid Mechanics 59~(04) (1973) 721--736.

\bibitem{Hammack2}
J.~L. Hammack, H.~Segur, The {K}orteweg-de {V}ries equation and water waves.
  {II}. {C}omparison with experiments, J. Fluid Mech. 65 (1974) 289--313.

\bibitem{Hammack3}
J.~L. Hammack, H.~Segur, The {K}orteweg-de {V}ries equation and water waves.
  {III}. {O}scillatory waves, J. Fluid Mech. 84~(2) (1978) 337--358.

\bibitem{TrogdonSOKdV}
T.~Trogdon, S.~Olver, B.~Deconinck,
  \href{http://www.sciencedirect.com/science/article/pii/S016727891200053X}{Nu%
merical inverse scattering for the {Korteweg-de Vries} and modified
  {Korteweg-de Vries} equations}, Physica D 241 (2012) 1003--1025.
\newblock \href {http://dx.doi.org/10.1016/j.physd.2012.02.016}
  {\path{doi:10.1016/j.physd.2012.02.016}}.
\newline\urlprefix\url{http://www.sciencedirect.com/science/article/pii/S01672%
7891200053X}

\bibitem{Deconinck-theta}
B.~Deconinck, M.~Heil, A.~Bobenko, M.~van Hoeij, M.~Schmies,
  \href{http://dx.doi.org/10.1090/S0025-5718-03-01609-0}{Computing {R}iemann
  theta functions}, Math. Comp. 73 (2004) 1417--1442.
\newblock \href {http://dx.doi.org/10.1090/S0025-5718-03-01609-0}
  {\path{doi:10.1090/S0025-5718-03-01609-0}}.
\newline\urlprefix\url{http://dx.doi.org/10.1090/S0025-5718-03-01609-0}

\bibitem{klein}
J.~Frauendiener, C.~Klein,
  \href{http://dx.doi.org/10.1007/s11005-006-0068-4}{Hyperelliptic
  theta-functions and spectral methods: {K}d{V} and {KP} solutions}, Lett.
  Math. Phys. 76 (2006) 249--267.
\newblock \href {http://dx.doi.org/10.1007/s11005-006-0068-4}
  {\path{doi:10.1007/s11005-006-0068-4}}.
\newline\urlprefix\url{http://dx.doi.org/10.1007/s11005-006-0068-4}

\bibitem{Lax}
P.~D. Lax, Periodic solutions of the {K}d{V} equation, Comm. Pure Appl. Math.
  28 (1975) 141--188.

\bibitem{TrogdonFiniteGenus}
T.~Trogdon, B.~Deconinck, A {Riemann--Hilbert} problem for the finite-genus
  solutions of the {KdV} equation and its numerical solution, to appear in
  Physica D.

\bibitem{TrogdonDressing}
T.~Trogdon, B.~Deconinck, A numerical dressing method for the superposition of
  solution of the {KdV} equation, submitted for publication.

\bibitem{teschl-finite-gap}
I.~Egorova, K.~Grunert, G.~Teschl,
  \href{http://dx.doi.org/10.1088/0951-7715/22/6/009}{On the {C}auchy problem
  for the {K}orteweg-de {V}ries equation with steplike finite-gap initial data.
  {I}. {S}chwartz-type perturbations}, Nonlinearity 22~(6) (2009) 1431--1457.
\newblock \href {http://dx.doi.org/10.1088/0951-7715/22/6/009}
  {\path{doi:10.1088/0951-7715/22/6/009}}.
\newline\urlprefix\url{http://dx.doi.org/10.1088/0951-7715/22/6/009}

\bibitem{teschl-asymptotics}
A.~Mikikits-Leitner, G.~Teschl,
  \href{http://dx.doi.org/10.1007/s11854-012-0005-7}{Long-time asymptotics of
  perturbed finite-gap {K}orteweg-de {V}ries solutions}, J. Anal. Math. 116
  (2012) 163--218.
\newblock \href {http://dx.doi.org/10.1007/s11854-012-0005-7}
  {\path{doi:10.1007/s11854-012-0005-7}}.
\newline\urlprefix\url{http://dx.doi.org/10.1007/s11854-012-0005-7}

\bibitem{McKean-Trubowitz}
H.~P. McKean, E.~Trubowitz,
  \href{http://dx.doi.org/10.1090/S0002-9904-1978-14542-X}{Hill's surfaces and
  their theta functions}, Bull. Amer. Math. Soc. 84 (1978) 1042--1085.
\newblock \href {http://dx.doi.org/10.1090/S0002-9904-1978-14542-X}
  {\path{doi:10.1090/S0002-9904-1978-14542-X}}.
\newline\urlprefix\url{http://dx.doi.org/10.1090/S0002-9904-1978-14542-X}

\bibitem{GGKM}
C.~S. Gardner, J.~M. Greene, M.~D. Kruskal, R.~M. Miura, Method for solving the
  {Korteweg--de Vries} equation, Phys. Rev. Lett. 19 (1967) 1095--1097.

\bibitem{FokasUnified}
A.~S. Fokas, A Unified Approach to Boundary Value Problems, SIAM, Philadelphia,
  PA, 2008.

\bibitem{Dressing}
E.~V. Doktorov, S.~B. Leble, A dressing method in mathematical physics, Vol.~28
  of Mathematical Physics Studies, Springer, Dordrecht, 2007.

\bibitem{ZakharovDressing}
V.~E. Zakharov, On the dressing method, in: Inverse methods in action
  ({M}ontpellier, 1989), Inverse Probl. Theoret. Imaging, Springer, Berlin,
  1990, pp. 602--623.

\bibitem{TrogdonPeriodic}
T.~Trogdon, B.~Deconinck,
  \href{http://dx.doi.org/10.1080/00036811.2010.549480}{The solution of linear
  constant-coefficient evolution {PDE}s with periodic boundary conditions},
  Appl. Anal. 91 (2012) 529--544.
\newblock \href {http://dx.doi.org/10.1080/00036811.2010.549480}
  {\path{doi:10.1080/00036811.2010.549480}}.
\newline\urlprefix\url{http://dx.doi.org/10.1080/00036811.2010.549480}

\bibitem{DLMF}
F.~W.~J. Olver, D.~W. Lozier, R.~F. Boisvert, C.~W. Clark, NIST Handbook of
  Mathematical Functions, Cambridge University Press, 2010.

\bibitem{FokasComplexVariables}
M.~J. Ablowitz, A.~S. Fokas, Complex Variables: Introduction and Applications,
  Cambridge University Press, 2003.

\bibitem{TrogdonSONLS}
T.~Trogdon, S.~Olver, Numerical inverse scattering for the focusing and
  defocusing {nonlinear Schr\"odinger} equations, Proc. R. Soc. A. 469.

\bibitem{TrogdonSONNSD}
S.~Olver, T.~Trogdon, Nonlinear steepest descent and the numerical solution of
  {Riemann--Hilbert} problems, to appear in Comm. Pure Appl. Math.

\bibitem{SORHFramework}
S.~Olver, \href{http://dx.doi.org/10.1007/s00211-012-0459-7}{A general
  framework for solving {R}iemann-{H}ilbert problems numerically}, Numer. Math.
  122~(2) (2012) 305--340.
\newblock \href {http://dx.doi.org/10.1007/s00211-012-0459-7}
  {\path{doi:10.1007/s00211-012-0459-7}}.
\newline\urlprefix\url{http://dx.doi.org/10.1007/s00211-012-0459-7}

\bibitem{SOPainleveII}
S.~Olver, Numerical solution of {Riemann--Hilbert} problems: {Painlev\'e II},
  Found. Comput. Math.

\end{thebibliography}







\end{document}